\documentclass[conference]{IEEEtran}
\IEEEoverridecommandlockouts
\usepackage{cite}
\usepackage{comment}
\usepackage{amsmath,amssymb,amsfonts}

\usepackage{algpseudocode}
\usepackage{graphicx}
\usepackage{subfigure}
\usepackage{float}
\usepackage{subfig}
\usepackage[font=footnotesize,labelfont=rm]{caption}
\usepackage{textcomp}
\usepackage{xcolor}
\usepackage{url}
\usepackage{multirow}
\usepackage{dblfloatfix}
\usepackage[linesnumbered,ruled,vlined]{algorithm2e}
\SetKwInput{KwInput}{Input}                
\SetKwInput{KwOutput}{Output}              

\def\BibTeX{{\rm B\kern-.05em{\sc i\kern-.025em b}\kern-.08em
    T\kern-.1667em\lower.7ex\hbox{E}\kern-.125emX}}

\begin{document}
\title{6G for Connected Sky: A Vision for Integrating Terrestrial and Non-Terrestrial Networks}

\author{\small {\bf Mustafa Ozger$^1$, Istvan Godor$^2$, Anders Nordlow$^3$, Thomas Heyn$^4$, Sreekrishna Pandi$^5$, Ian Peterson$^6$, Alberto Viseras$^7$,} \\ \small {\bf Jaroslav Holis$^8$, Christian Raffelsberger$^9$, Andreas Kercek$^9$, Bengt M\"olleryd$^{10}$,  Laszlo Toka$^{11, 12}$, Gergely Biczok$^{11, 12}$,}\\ \small {\bf Robby de Candido$^{13}$, Felix Laimer$^{14}$,  Udo Tarmann$^{15}$, Dominic Schupke$^6$, and Cicek Cavdar$^1$} \\ 
\small $^1$KTH Royal Institute of Technology, Sweden $^2$Ericsson, Hungary  $^3$Ericsson, Sweden  $^4$Fraunhofer IIS, Germany $^5$Meshmerize, Germany \\ \small  $^6$Airbus, Germany $^7$Motius, Germany $^8$Deutsche Telekom, Germany  $^9$Lakeside Labs, Austria $^{10}$PTS, Sweden \small$^{11}$AITIA, Hungary \\ $^{12}$Budapest Univ. of Technology and Economics, Hungary $^{13}$Skysense, Sweden $^{14}$Bernard-Gruppe, Austria $^{15}$LCA, Austria}
\maketitle

\begin{abstract}
In this paper, we present the vision of our project 6G for Connected Sky (6G-SKY) to integrate terrestrial networks (TNs) and non-terrestrial networks (NTNs) and outline the current research activities in 6G research projects in comparison with our project. From the perspectives of industry and academia, we identify key use case segments connecting both aerial and ground users with our 6G-SKY multi-layer network architecture. We explain functional views of our holistic 6G-SKY architecture addressing the heterogeneity of aerial and space platforms. Architecture elements and communication links are identified. We discuss 6G-SKY network design and management functionalities by considering a set of inherent challenges posed by the multi-layer 3-dimensional networks, which we termed as combined airspace and NTN (combined ASN). Finally, we investigate additional research challenges for 6G-SKY project targets.
\end{abstract}

\vspace{-2mm}
\section{Introduction}\label{sec.intro}

Digital airspace is transforming rapidly with major societal change programs such as EU SES and SESAR where Europe aims to create European Digital Sky by 2040 \cite{sesar}. These programs have increased the need for connectivity, sustainability, a higher degree of autonomous operations, automation of Air Traffic Management (ATM)/Universal Traffic Management (UTM), and new business models. 6G will play a crucial role as a key technology enabler to realize digital airspace.  

Advancements in aviation and space industries have increased the number of flying vehicles (FVs) at different altitudes in the sky. For instance, unmanned aerial vehicles (UAVs) and flying taxis are foreseen as new vehicles in the sky. Furthermore, an increasing demand is observed for air travel and cargo. There are also intensive efforts to deploy new satellite constellations. Another segment of FVs is high altitude platforms (HAPS) and HAP stations as  International Mobile Telecommunications (IMT) base stations (HIBS), which have received significant interest from telecom operators to complement their terrestrial networks (TNs). 

3D network architectures combining terrestrial, HAPS/HIBS and satellite networks have the full potential to provide ubiquitous broadband connectivity with a managed latency to both terrestrial and aerial users. Novel methods must be studied for 6G to secure higher spectrum utilization efficiency and network operation with lower power consumption by exploiting solar power and green hydrogen. Connected sky complementing 6G terrestrial network will bring required data service for everyone everywhere. 

There are increasing efforts from industry and academia via research projects to design next-generation networks. 5D-AeroSafe project \cite{T1} develops UAV-based services to ensure the safety and security of airports and waterways. ETHER \cite{T21} focuses on a 3D multi-layered architecture with a design focus on the antenna, waveform, handover and network management. 6G-NTN \cite{T22} integrates NTN components in 6G to deliver ultra-reliable low latency communication (URLLC) and enhanced mobile broadband (eMBB), emergency and high-accuracy location services. Hexa-X \cite{T2} proposes new radio access technologies at high frequencies, high-resolution localization, and intelligent network design. DEDICAT6G \cite{T3} develops 6G for human-centric applications. 6G BRAINS \cite{T16} develops AI-driven multi-agent deep reinforcement learning solutions for resource allocation in machine-type communication networks for future industrial networks. AI@EDGE \cite{T17} aims at developing secure and trustworthy AI solutions to automatically manage heterogeneous mobile edge computing resources. MARSAL \cite{T18} aims at developing a complete framework to manage and orchestrate network resources in beyond 5G networks with optical and wireless infrastructure. DAEMON \cite{T19} develops novel approaches for network intelligence design to enable high-performance, sustainable and extremely reliable zero-touch network systems. REINDEER \cite{T20} aims at achieving perceived zero latency and uninterrupted availability in time and location via developing smart communication technologies. 

\begin{table*}[]
\vspace{2mm}
\centering
\caption{Use case segments for all aerial platforms on all altitude levels, ground and space.}
\begin{tabular}{|ll|l|l|l|}
\hline
\multicolumn{2}{|l|}{\textbf{Connectivity Cases}} &
  \textbf{\#} &
  \textbf{Use Case Segments} &
  \textbf{Scenarios} \\ \hline
\multicolumn{2}{|l|}{{\begin{tabular}[c]{@{}l@{}}Airspace \\ communication\end{tabular}}} &
  1 &
  Commercial Airplane Traffic &
  \begin{tabular}[c]{@{}l@{}}Passenger operation (inflight entertainment connectivity)\\ Aircraft operation (communication between ATC and pilots)\\ Crew operation and ground operation\end{tabular} \\ \cline{3-5} 
\multicolumn{2}{|l|}{} &
  2 &
  Urban Air Mobility (UAM) &
  Smart transportation (flying taxis), smart city services, smart logistics \\ \cline{3-5} 
\multicolumn{2}{|l|}{} &
  3 &
  Verticals beyond UAM &
  \begin{tabular}[c]{@{}l@{}}Public safety, peacekeeping, and defense applications\\ through LOS/BVLOS operation of FVs\end{tabular} \\ \cline{3-5} 
\multicolumn{2}{|l|}{} &
  4 &
  \begin{tabular}[c]{@{}l@{}}Digital Airspace and TN integration\\  towards management systems\end{tabular} &
  Air traffic management (ATM), UTM, national security  systems \\ \hline
\multicolumn{1}{|l|}{\multirow{3}{*}{NTN}} &
  \multirow{2}{*}{\begin{tabular}[c]{@{}l@{}}Capacity \\ \\ Extension\end{tabular}} &
  5 &
  \begin{tabular}[c]{@{}l@{}}Connectivity for rural and \\ remote areas\end{tabular} &
  \begin{tabular}[c]{@{}l@{}}Extending 3GPP services to rural and remote areas\\ Support of URLLC services, including HAPS/HIBSs integration\end{tabular} \\ \cline{3-5} 
\multicolumn{1}{|l|}{} &
   &
  6 &
  \begin{tabular}[c]{@{}l@{}}Satellite/HAPS/HIBS for backhauling\\  - non-3GPP services\end{tabular} &
  \begin{tabular}[c]{@{}l@{}}Satellite and HAPS platforms for earth observation, disaster \\ management and critical infrastructure monitoring\end{tabular} \\ \cline{2-5} 
\multicolumn{1}{|l|}{} &
  \begin{tabular}[c]{@{}l@{}}Coverage \\ Extension\end{tabular} &
  7 &
  Internet of Things &
  \begin{tabular}[c]{@{}l@{}}power efficient communication ranging from logistical tracking, \\ telemetry, remote monitoring, geo fencing, security, etc.\end{tabular} \\ \hline
\end{tabular}
\label{tab:use_case}
\vspace{-4mm}
\end{table*}

\begin{figure*}[th!]
 	\centering
        \includegraphics[width=0.85\textwidth]{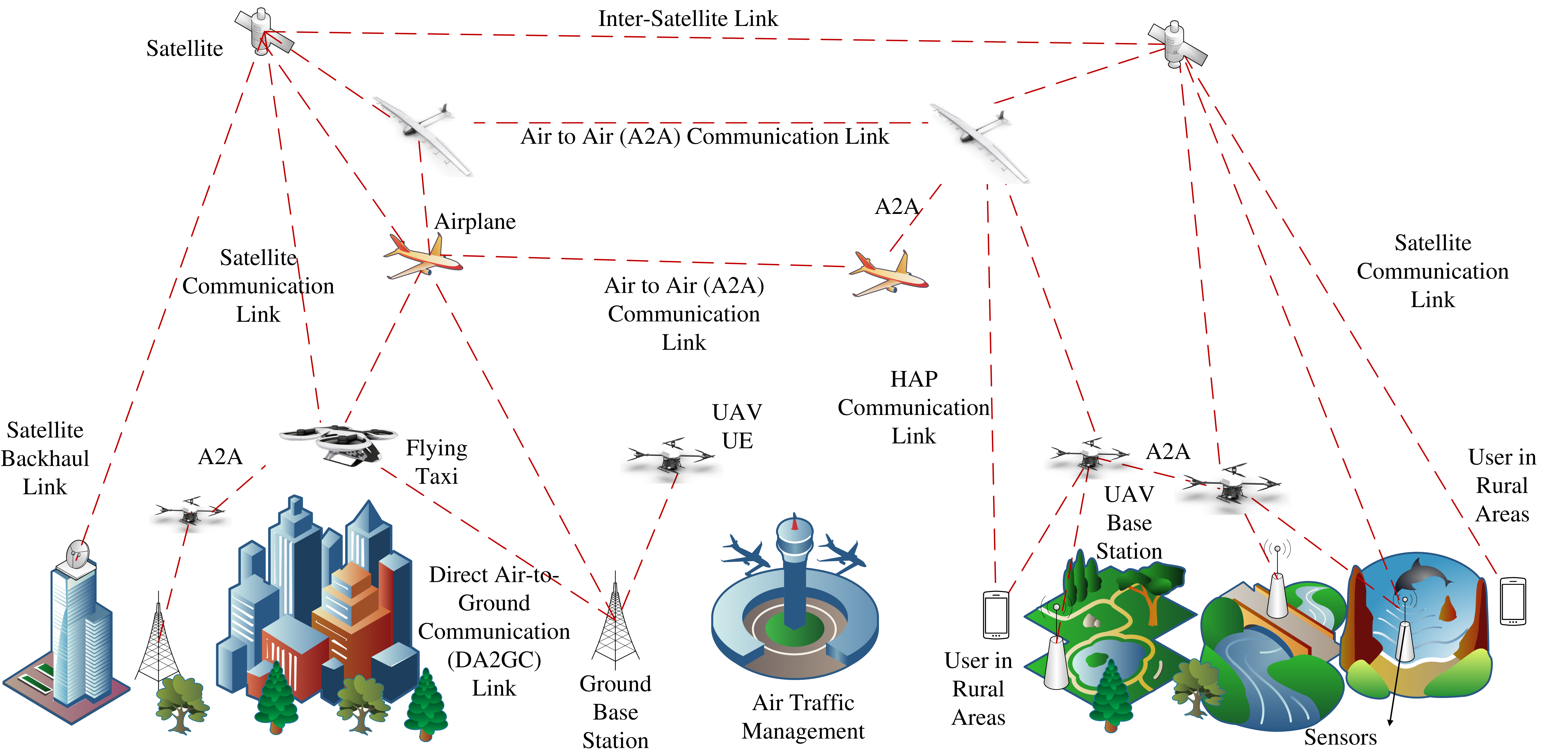}
 	\caption{Combined airspace and NTN (combined ASN) for 6G limitless connectivity to all aerial platforms and terrestrial users.}
 	\label{fig:connect}
 \end{figure*}

Despite notable efforts in the mentioned projects, holistic designs of 6G wireless networks integrating TNs and NTNs to provide connectivity to both ground and aerial users are absent. Most of these projects focus on designing machine learning (ML)-driven solutions to provide energy efficiency, zero perceived delay, network management and orchestration for beyond 5G and 6G networks. However, the primary focus of most of them is TNs disregarding FVs and satellites. On the other hand, ETHER and 6G-NTN projects focus on integrating NTN elements for certain use cases such as connecting underserved areas. Furthermore, Hexa-X develops 6G radio technologies for ground users’ communication. In our project 6G for Connected Sky (6G-SKY) \cite{6g-sky}, we design a holistic adaptive AI and cloud-native network architecture to unlock potentials of 3D communications with a focus on diverse use cases such as urban air mobility (UAM) and also regulations on the spectrum. The main difference between the mentioned projects and the 6G-SKY project is that it enables reliable and robust connectivity for aerial and ground users via flexible and adaptive network architectures adopting multiple technologies such as satellite and direct air-to-ground communication (DA2GC). Furthermore, novel radio technologies will be proposed to support high capacity, reliable and secure DA2GC and air-to-air communication (A2AC) links as well as low delay and reliable command and control links. The devised novel radio technologies and resource allocation schemes will address challenges posed by wireless communication and networking and regulations with respect to safety, airspace management, and frequency usage. As compared to the other mentioned projects, the 6G-SKY project extends the architecture scope into the third dimension together with a holistic and integrated network architecture to support a diverse set of quality of services (QoS) for both aerial and ground users. The 6G-SKY project will take into account different characteristics of FVs and exploit UAVs, HAPS/HIBS, and satellites for connectivity. Fig. \ref{fig:connect} illustrates our holistic architecture. We also propose a new term: combined airspace and NTNs, abbreviated as combined ASN to name our holistic network architecture. This architecture extends the terrestrial cellular architecture through communication services that can be provided to and/or by flying platforms in airspace such as airplanes, HAPs, and UAVs - including DA2GC - and by satellites. 

The multi-layered architecture in 6G-SKY necessitates the joint interplay between the telecommunication, aviation, and space industries. The project innovations address key elements, to demonstrate 5G Advanced and 6G features that underpin the architecture, to de-risk technical challenges at an early stage, and to ensure corresponding interoperability standards, commencing in 3GPP in $\sim$5 years for 6G. 
De-risking addresses principal showstoppers such as the use of spectrum between ground/air/space entities and technical issues such as communications performance of mobile ground/air/space entities to eventually provide 6G services. 

\vspace{-1mm}
\section{Use Cases and Scenarios}
\label{sec:usecases}
6G-SKY operates in a highly dynamic environment with many use cases and services to be provided. 6G-SKY connectivity services are divided into two categories. The first one is \textit{airspace communication} referring to connecting FVs via DA2GC, A2AC, etc. This service category covers data link and command and control (C2) link services for FVs. The other service category is \textit{NTN connectivity service} to connect terrestrial users. This connectivity service is for either \textit{capacity extension} or \textit{coverage extension}. To cater to all these use cases, use case segments have been defined to cover attached services. The use case segments are defined from a commercial perspective as seen in Table \ref{tab:use_case} with connectivity services.


A number of use case segments are identified under airspace communications connectivity for FVs. \textit{Segment 1} captures the vision of communication towards having the same user experience on an airplane as on the ground level, which includes different types of operations in airplanes such as passenger communication. \textit{Segment 2} aims to enable UAM that will impact 6G innovation through smart transportation, cities and logistics via TN and NTN integration forming a digital society. The connectivity services we target in \textit{Segment 2} are data and C2 link services for FVs. \textit{Segment 3} captures the use cases beyond UAM such as public safety applications with beyond visual line of sight (BVLOS) operation. As this segment aims to connect FVs, the main connectivity services are data and C2 communication services.  \textit{Segment 4} interacts with all aerial platforms and different management systems such as UTM, in which the data and C2 services are key to digitalizing airspace. On the other hand, certain use case segments capture NTN connectivity for terrestrial users. \textit{Segment 5} provides the same level of network performance in rural/remote areas as in densely populated areas with 3GPP services to extend ground communication capacity. \textit{Segment 6} is for use cases employing wireless networks as a connection point and transport of satellite/HAPS/HIBS-related data traffic for non-3GPP services. This segment is for the capacity extension for ground users. \textit{Segment 7} is mainly for IoT connectivity to support power-efficient communication, where the attached services are coverage extensions.  

\begin{figure}[b!]
 	\centering
 	\includegraphics[width=0.84\columnwidth]{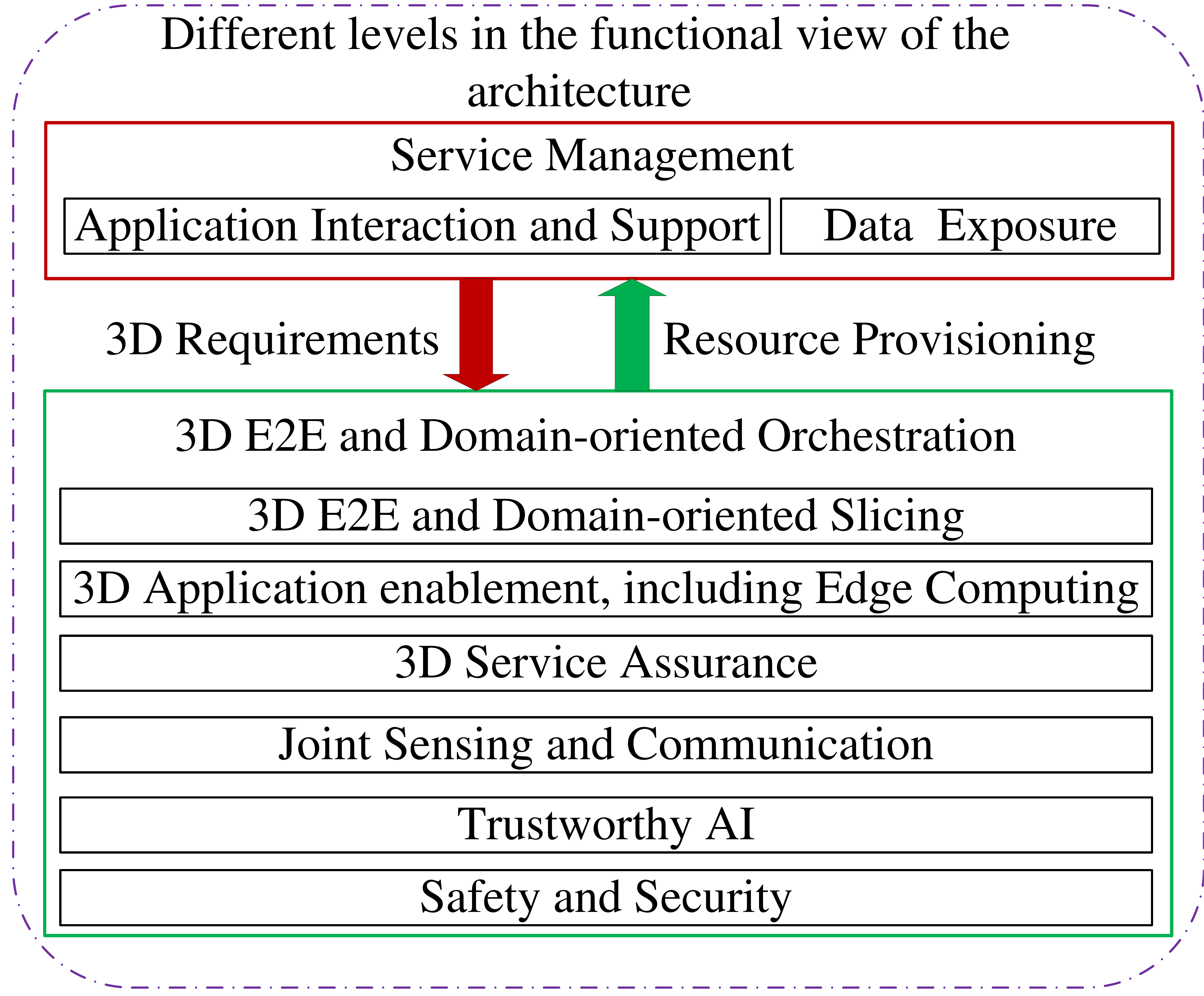}
 	\caption{6G-SKY architecture functional view.}
 	\label{fig:architecture}
\vspace{-4mm}
 \end{figure}

\vspace{-1mm}
\section{6G-SKY Architecture}
\label{sec:arch}
To provide limitless connectivity to all aerial platforms at all altitude levels and users on the ground, the 6G-SKY network architecture will impact all functional levels of 3D wireless systems. Fig. \ref{fig:architecture} presents different functionality to stress the need for this integration and architecture functions from a 3D perspective. The main considerations of this architecture are: all wireless concepts must be reviewed from a 3D perspective, all functional levels such as service management and network orchestration must be considered, communication options increase both for digital airspace and TN for same use cases, and there is a need for the architecture to support both legacy standards (4G, 5G, 5G Advanced) and new standards (6G). Based on the different levels in the functional view of our architecture, the service management function activates, changes and closes a service from a service life cycle perspective. It is also a borderline for external service activation across wireless systems. Service management interfaces customer intent, service level specifications and key performance indicators (KPIs) with use case segments. To this end, application interaction and support functions are for wireless systems activities related to service management and support requests from the application layer. Data exposure makes the network data available to extend services. These services drive 3D requirements from the network. The function interacting with the service management is 3D end-to-end (E2E) and domain-oriented orchestration to provision resources according to the requirements. This function includes slicing to guarantee required services for provisioning. Other functions in the 6G-SKY architecture are joint sensing and communication, edge  computing, trustworthy AI, safety and security. 

Each of NTNs components such as satellite and HAPS/ HIBS is operated and designed independently mostly in competition with one another to provide connectivity to both terrestrial and aerial users. TNs up to 3GPP Release 16 are also designed independently to provide connectivity and coverage for different connectivity services resulting in the over-provisioning of resources with very high deployment costs. They all have different pros and cons, with no single technology yet that fits to support a large diversity of requirements raised by aerial and terrestrial users. However, the first time that NTN networks are supported in 3GPP is the current Release 17. 6G-SKY project aims to continue and extend this approach and design the integration of different NTNs with TNs combining the strengths of each technology in a joint solution minimizing the cost and satisfying a diverse set of QoS requirements from terrestrial and aerial users. The 6G-SKY project focuses on developing an autonomously adaptive ``5G and beyond" network in 3D to shape itself considering the heterogeneity of the following three important aspects:

\textbf{Communication links:} Satellite communication links (different delay and throughput options depending on the satellite technology); DA2GC links between the ground BSs and aerial users; A2AC links among same and different types of aerial vehicles in the 3D space; HAPS communication links.

\textbf{Aerial users with different QoS requirements:} UAVs have different throughput, delay and reliability requirements depending on use cases. BVLOS operations, swarm coordination, etc., require ultra-reliable and robust connectivity, while passenger communications in commercial airplanes require high-capacity links. Although UAVs act as flying sensors (for inspection, traffic control, etc.), requiring massive payload data often high-capacity links beside low latency QoS for UAV coordination. UAVs (in single, teaming, or swarming mode) may themselves be part of the communication infrastructure among aerial and terrestrial users. In this case, the UAVs act as relays where swarm coordination has to optimize e.g., certain communication coverage and QoS requirements of the target application. In this way, a highly dynamic and adaptive ad-hoc network can be realized for particular applications like a temporary enhanced communication need in critical traffic situations, autonomous truck platooning, and emergency situations. Again, low latency, as well as high-capacity links, are required at the same time. In addition, flying taxis as elements of UAM impose a new set of requirements to satisfy their safe and secure operations similar to Air Traffic Management (ATM). URLLC and enhanced mobile broadband (eMBB) communications are enablers of these applications.

\textbf{Terrestrial users located at regions hard to access:} Sensors located in such regions may require massive machine-type communication (mMTC) coverage. On the other hand, ground users require high throughput connectivity such as eMBB. Industrial applications or future person and goods mobility based on automated vehicles and their tight coupling to an intelligent infrastructure with related services (emergency- traffic-, information, route planning, etc.) in remote areas with critical machine-type communications (cMTC) may require URLLC. This, for instance, will particularly be the case when the level of automation in future mobility (goods and personal) rises and requires machine-type communication (V2V-Vehicle-to-Vehicle and V2I-Vehicle-to-Infrastructure). This will be supported by both, ground-based and aerial communication infrastructure as proposed in this project.

To address the heterogeneity in different aspects of the 6G-SKY project, the first objective is to build an integrated 3D network, i.e., combined ASN, as seen in Fig \ref{fig:connect}. It combines distinct communication technologies such as DA2GC, satellite communication, and A2AC to satisfy the connectivity requirements of both aerial and ground users. It should be also robust to the changes in the networks such as variations in link capacities via utilizing multiple communication technologies. It supports constructing flexible network topologies to adapt network densities and traffic variations for numerous use cases and scenarios via complying with spectrum and airspace regulations. It aims at providing the safety and security of FVs via newly designed solutions such as sense and avoid mechanisms. To the best of our knowledge, there is no effort to provide a holistic network architecture design as in Fig. \ref{fig:connect}, which addresses the heterogeneity of the FVs, communication technologies, and communication requirements of different users with machine learning tools for adaptive and robust communications. Additionally, a special consideration addresses the exploitation of the new integrated sensing capabilities of 6G within the architecture for network operations and sensing services.

Due to the heterogeneous requirements of both ground and aerial users, the second objective is to design connectivity links. Commercial airplanes at high altitudes require high-capacity links. UAVs require URLLC for BVLOS and swarming applications. Also, UAVs require broadband connectivity for their reconnaissance or inspection applications or when acting as communication relays within the 6G-SKY architecture. Flying taxis have highly reliable connectivity to avoid any casualties. To satisfy the diverse requirements of aerial users, 6G communication technologies, which are essentially unexplored in the NTN context, will be utilized such as Terahertz communication, millimeter wave communications and reflective intelligent surfaces. Satellite, DA2GC and HAPS links are the main connectivity options as seen in Fig. \ref{fig:connect} to address the connectivity demands. Our objective is not limited to the design of connectivity links for aerial users. We also target to provide connectivity to ground users in rural areas via satellite, HAPS, and even UAV technologies. The latter can provide (temporarily) enhanced connectivity for (personal and goods) transport in rural areas. This will especially be the case for the improvement of V2V and V2I connectivity when automated vehicles and intelligent infrastructure will emerge. This also requires a particular swarm coordination of UAVs which we will develop in this project. Moreover, UAVs can serve as flying eyes to provide infrastructure and vehicles with up-to-date traffic and emergency information. The use cases for the ground users are mMTC, cMTC and eMBB. 

The third objective is the integration of multiple links in 3D from multiple technologies to design 3D networks for aerial and ground users. To this end, we will propose novel solutions with adaptive network topology shaping to respond to the changes in the 3D environment. Cell-free network design to provide adaptive user-centric network formation as an enabling technology of 6G networks will be crucial to integrate multiple frequencies and different communication technologies. This aims at autonomous control of handovers among different network technologies as flying UEs move in 3D space as well as coordinate them accordingly (e.g. coordination of UAVs based on self-organization and swarm intelligence). Due to the use of high frequency, beam tracking and beamforming algorithms will be proposed, which require precise localization of FVs in 3D space. For ground users in hard-to-access areas such as rural areas, NTNs can provide connectivity and coverage when terrestrial networks are not capable to meet different connectivity requirements. Covering these ground users with certain types of communications such as eMBB for video transmissions, sensor data, URLLC for critical IoT applications and extended coverage for massive MTC applications are possible via networking solutions through satellites, UAVs, and HAPS. 


    
    

\section{6G-SKY Network Design}
\label{sec:netdesign}
\subsection{Network Design and Management Functionality} 

Despite the wide level of heterogeneity of current communication network services, the communication needs of various types of users being either humans, machines, or even larger infrastructures, current networks mainly consider a flat, 2D type network architecture. These flat networks focus on BSs  serving terrestrial terminals in mobile networks or satellite-based communication for some selected groups including aerial users like airplanes as well as terrestrial users on the ground like smartphones and stationary terminals.

An emerging challenge is to complement existing TNs forming a flexible 6G hybrid architecture integrating terrestrial, HAPS and satellite layers with an additional aerial layer with users such as UAVs. As in terrestrial communications, the service requirements are diverse in aerial applications and dynamically changing such that a combined ASN design should be able to adapt its topology and resources according to the needs of both terrestrial and aerial users. 

The main focus of radio resource management in the combined ASN is to control and harmonize the network side domain towards the optimal resource allocation and load balancing between the connection domains including the cost, QoS demands as well as energy efficiency aspects of connections. Moreover, such multi-domain networks might be operated by different operators, so multiple operator policies have to be mapped to the service control logic. As the network complexity increases, when terrestrial and aerial users are served by a multi-layer network architecture, the optimization of the radio resource management by AI is advantageous.

In the multi-layer 3D network architecture, the users might be connected to multiple serving layers (e.g., dual or multi-connectivity to traditional terrestrial networks and satellites). Hence, the mobility handling (handover) depends not only on their altitude, but special attention has to be taken to the typical LOS conditions of aerial users and the resulting potential interference between aerial and terrestrial communication links as well as largely overlapping service areas (aka cells in traditional terrestrial communications). Thus, mobility management greatly relies on the accurate reported or predicted localization information of the UEs and knowledge of the individual link qualities. 

Localization should be a 6G service for users independently from the instantaneous availability of connection domains. Not only the networks but also each UE device can be considered as a sensor. Inherent possibilities of joint communication and sensing  technologies, beam-forming, and ultra-wideband scanning technologies available to precisely locate a given UE, especially in LOS environments are envisaged to be typical for aerial users in the future, as well.
\subsection{Aerial Mesh Networks}

One of the novel elements in the combined ASN is the incorporation of direct D2D communication, specifically A2AC, into the 6G architecture. The motivation is to extend the network coverage even further in places where no other means of communication are available for the users. However, relying solely on A2AC links for network coverage and use cases such as drone swarms would be inadequate as it is limited to a single hop of communication. To address this issue, the 6G-SKY project aims to incorporate a managed multi-hop network, commonly referred to as a mesh network, to combine multiple A2AC links into a single network. The mesh network should also seamlessly interface with other parts of the heterogeneous network architecture, such as terrestrial, HAPS/HIBS, and satellite links, in a way that is transparent to the user equipment, allowing for seamless connectivity and efficient use of resources. The transparency of the overall system considers that only a limited amount of communication agents may possess interfaces to other systems and serve as seamless gateways between the networks.

One of the key use cases that an aerial mesh network will enable is the operation of UAV swarms. A heterogeneous network must be designed that incorporates mesh networks to support this scenario. Traditional mesh networks are optimized for maximizing throughput, but for UAV swarms, a different approach is needed to prioritize resilience and low latency for command and control data. This is critical for modern precise distributed swarm control algorithms, which rely on real-time communication between the UAVs. Furthermore, the development of more sophisticated swarming algorithms is needed to adaptively react to the changing radio environment.

The mesh network developed should not only utilize the multi-hop aspect but also leverage multi-path to transport high-priority data for improved latency and reliability performance. Another approach is the utilization of network slicing within the mesh network, allowing for different traffic streams to have distinct QoS specifications that can be centrally managed in the control plane. This will enable the network to adapt and optimize the resources for different use cases and traffic types, thus providing a more efficient and reliable network.

\section{Additional Challenges Addressed in 6G-SKY}
\label{sec:open}

\textbf{Spectrum:} ITU regulates the general use of frequencies in the world through its Radio Regulations and has overall international responsibility for satellite network coordination, notification as well as bringing into use aspects. Hence, ITU determines the allocation of spectrum available for satellites. Satellite components are in space and are therefore outside the jurisdiction of national regulatory authorities (NRA), like PTS in Sweden. However, assignments and licensing of spectrum for communication to satellites from Earth stations are decided by NRAs. 

At World Radio Conferences (WRC), regulatory advice regarding spectrum sharing has been agreed upon, in which recommendations are to maintain a minimum separation angle as well as guidelines covering maximum transmitted powers for different bands and limits of power-flux densities (PFD) from space stations. More research is required, to verify the coordination, sharing and compatibility between new and old technologies supported at different altitudes and orbits and used in the same or adjacent frequency bands.

ITU regulations, the only spectrum band where HAPS can currently act as a cellular base station is $2.1$ GHz. However, WRC-23 agenda item 1.4 is looking to consider HAPS mobile services in certain frequency bands already identified for IMT: $694-960$ MHz; $1710-1885$ MHz and $2500-2690$ MHz. 6G-SKY project aims to establish a path toward sustainable 6G that results in higher energy and spectral efficiency compared to previous mobile network generations. 




\textbf{Security and Safety:} From the safety aspect, the greatest challenge regarding safe urban airspace operations is the coordination of flight missions comprising both manned and unmanned aircraft. To this end, a complete single-source picture of the sky is indispensable. Coming U-space \cite{u-space} and FAA \cite{faa} regulations enforcing geofencing will prevent UAVs from flying unintentionally at restricted locations, but those regulations will not stop non-compliant UAVs and pilots with malicious intent from entering restricted or controlled airspace. Detecting and locating UAVs not broadcasting or providing their location (non-cooperative) is required to ensure a complete, single-source picture of the sky in urban areas. Another challenge in this context is then how to convey, in real-time, to manned aircraft that UAVs are present in the airspace. To address the information deficit among airspace users, we foresee the emergence of drone detection and positioning systems and ground-based low-power Automatic Dependent Surveillance–Broadcast (ADS-B) transmitters and ADS-B receivers.

From the security aspect, the proposed novel holistic architecture, making use of heterogeneous links, comes with its own security requirements. On top of the inherent challenges in wireless security, we foresee that thwarting spoofing attacks will become a focus point of related security efforts. Such attacks could potentially cripple two major functionalities of the novel architecture utilizing unauthenticated communication protocols: i) advanced localization (via global navigation satellite system (GNSS) spoofing) and ii) the above-mentioned drone detection (via ADS-B spoofing). In addition, UAVs introduce an entirely new set of security challenges: they can be operated either by remote control or autonomously using onboard computers; accordingly, the UAV system is vulnerable to attacks that target either the cyber and/or physical elements, the interface between them, the wireless link, or even a combination of multiple components.

\textbf{Trustworthy AI:}  Many industries including aviation consider AI in critical operational contexts \cite{easa}. AI can act on different layers such as physical and network layers in 6G. On one hand, aviation can potentially profit from AI-optimized 6G networks to achieve performance parameters such as transmission reliability and data rate. On the other hand, implications for 6G services by AI-based networks that are used for critical aviation tasks need to be risk assessed. If possible, the network itself ensures the fulfillment of service parameters (e.g., defined in a service level agreement (SLA)). A network fails in fulfillment of service parameters (e.g., a longer disconnect during a flight), hence, trustworthy AI becomes critical in the 6G-SKY domain. For example, if our cases are based on time-series data, we can re-use some of the methods used in another use case that deals with this kind of data as well. This justifies the need of developing a Trustworthy AI platform that allows us to evaluate and develop AI models that can be trusted to be taken into the real world.  

\section{Conclusion}
\label{sec:con}
In this paper, we present an overview of the 6G-SKY vision to integrate terrestrial and non-terrestrial networks (NTNs). {We propose a new terminology as combined airspace and NTN (combined ASN) for 6G-SKY multi-layer networks.} Our project 6G-SKY will lay foundations of a holistic architecture adopting multiple technologies such as satellite and direct air-to-ground communication to realize different use case segments. We outline network design and management challenges in the combined ASN such as diverse service requirements, dynamically changing network topology, and heterogeneous network elements in the sky and on the ground with a combination of industrial and academic perspectives. Additional challenges such as spectrum regulations in 3D environments, trustworthy AI algorithms in aviation applications and the safety and security of flying vehicles are the distinctive aspects of the 6G-SKY project. 

\section*{Acknowledgment}
This work was supported in part by the CELTIC-NEXT Project, 6G for Connected Sky (6G-SKY), with funding received from the Federal Ministry for Economic Affairs and Climate Action under the contract number 01MJ22010B, Vinnova, Swedish Innovation Agency, the Austrian Federal Ministry for Climate Action, Environment, Energy, Mobility Innovation and Technology via the Austrian Research Promotion Agency (FFG) and Hungarian National Research, Development and Innovation Office, under the agreement no. 2020-1.2.3-EUREKA-2021-000006. The views expressed herein can in no way be taken to reflect the official opinion of the German ministry.

\end{document}